\documentclass[useAMS,usenatbib,dcolumn,usegraphicx,twocolumn,10pt]{mnarxiv}
\usepackage{times} 
\usepackage{placeins}
\newcommand{\ud}[1]{\mathrm{d}#1}

\newcommand{\br}[1]{\left(#1\right)}

\renewcommand{\sq}[1]{\left[#1\right]}
\newcommand{\eqref}[1]{(\ref{#1})}
\newcommand{\mpc}{\mathrm{Mpc}}
\newcommand{\kpc}{\mathrm{kpc}}

\newcommand{\msun}{\mathrm{M}_{\sun}}
\newcommand{\gu}{\mathrm{\ km\ s^{-1}\ kpc^{-1}}}

\title[Vertical gradients of azimuthal velocity in spiral galaxies]{Vertical gradients of azimuthal velocity in a global thin disk model of spiral galaxies NGC 2403, NGC 4559, NGC 4302 and NGC 5775.}
\author[]{Joanna Ja{\l}ocha$^{1}$,
{{\L}ukasz Bratek$^{1}$},
{Marek Kutschera$^{1,2}$,}
{Piotr Skindzier$^{2}$}
\\
$^{1}$Institute of Nuclear Physics,
Polish Academy of Sciences, Radzikowskego 152, PL-31342 Krak\'{o}w, Poland\\
$^{2}$Institute of Physics, Jagellonian University,  Reymonta 4, PL-30059 Krak\'{o}w, Poland}

\begin{document}

\date{\today}
\pagerange{\pageref{firstpage}--\pageref{lastpage}} \pubyear{2009}

\maketitle

\begin{abstract}
We estimate the vertical gradient of rotational velocity for several spiral galaxies in the framework of a global thin-disc model, using the approximation of quasi-circular orbits. We obtain gradients having a broad range of values, in agreement with measurements, for galaxies with both low and high gradients. To model the gradient, it suffices to know the rotation curve only. We illustrate, using the example of galaxy NGC 4302 with particularly high gradients, that mass models of galactic rotation curves that assume a significant spheroidal mass component reduce the predicted gradient value, which may suggest that the mass distribution is dominated by a flattened disc-like component. We conclude that the value and behaviour of the vertical gradient in rotational velocity can be used to study the mass distribution in spiral galaxies.\\

\medskip
\hrule
\flushleft \textbf{The definitive version is available at \\ \texttt{http://onlinelibrary.wiley.com/\\doi/10.1111/j.1365-2966.2010.17906.x/abstract}}
\medskip
\hrule

\end{abstract}


\section{Introduction}
Rotation curves carry important, though very limited, information about the kinematics of spiral galaxies. Until recently, rotation was determined mainly for matter moving in the direct neighbourhood of the galactic mid-plane. However, thanks to the constantly improving quality of the measurements, for several galaxies it has become possible to determine the vertical structure of the rotation above the galactic disc
(\citealt{bib:Fraternali2403}; \citealt{bib:Barbieri4559};  \citealt{bib:Heald5775},b; \citealt{bib:Heald4302}; \citealt{bib:Oosterloo}; \citealt{bib:gradients}).
This enables researchers to examine the mass distribution in spiral galaxies more accurately. The observations reveal that the azimuthal component of the galactic rotation falls off with the vertical distance from the galactic mid-plane. The direction of the fall-off distinguished may indicate that the galaxy dynamics is governed mainly by the gravity of masses concentrated close to the galactic mid-plane, forming a flattened, disc-like object. As for the high gradient values, one can hypothesize that the flattened galaxy component should comprise the gross dynamical mass of the galaxy.

Having this phenomenology in mind, we decided in our previous paper \citep{bib:MNRASgradient} to predict the gradient value in the Milky Way Galaxy. A flattened mass component was represented in our model by a thin disc, while the spheroidal mass components were represented by a spherically symmetric model. We analysed how the gradient estimates depended on the relative dynamical masses ascribed to these components. We observed that both the value and the behaviour of the predicted fall-off were strongly dependent on the assumed mass model. We obtained correct gradient values when the gross dynamical mass was attributed to the disc component, in agreement with our hypothesis. We also found that similar agreement with gradient measurements was possible in the global disc model of galaxy NGC 891. We therefore decided in the current work to carry out similar estimates for other four galaxies (NGC 2403, 4559, 5775 and 4302) in which the rotation structure has been determined above the galactic mid-plane.

In the current work we show that the thin-disc model, despite its simplicity, predicts correct values for the vertical gradient of rotational velocity, in agreement with the gradient measurements in the galaxies examined. In particular, this model predicts high gradient values that are difficult to explain by alternative gradient models (for example, entirely ballistic models of disc-halo flow give much lower values than observed; see \citet{bib:Heald4302}).

Furthermore, the different gradient behaviour predicted by various mass models for the same rotation curve offers one the opportunity to test the qualitative properties of the mass distribution in spiral galaxies. High gradient values imply a flattened mass distribution rather than a spheroidal one. This is particularly evident when the gradient is measured close to the mid-plane and beyond the centre, further away from the galactic bulge. In this region the shape of the gravitational field is determined mainly by the properties of the mass distribution in the galactic disc and halo, while the structure of the bulge is not very important there, owing to the fact that the contribution from the higher gravitational multipoles of a nearly spherically symmetric bulge falls off quickly with distance from the centre and the bulge can effectively be treated there as a point mass located in the centre. A qualitative argument for the suggestion that flattening of the mass distribution increases the gradient is provided by the following reasoning. In the case of a pure thin disc, one is more likely to obtain high gradient values in the mid-plane vicinity than in the opposite case of a purely spherical mass distribution. The reason is that, for the same rotation curve, the predicted gradient under the assumption of spherical symmetry is zero at the mid-plane, whereas it is non-zero in the disc model. Consequently, for a disc-like (i.e. mainly flattened) mass distribution, it is easier to produce high gradient values close to the disc plane, whereas a feature of a nearly spherically symmetric mass distribution is that close to the galactic plane its contribution to the overall gradient value is low. Thus, when the mass model is changed so as to reduce the disc mass component at the cost of increasing the contribution from the spheroidal mass component (making the overall mass distribution more spheroidal rather than more flattened and keeping sufficient the amount of mass required for the observed rotation velocity), the predicted gradient value will have been decreased, especially at larger radii (where the mass distribution becomes in this case even more dominated by the spheroidal component) and closer to the mid-plane. Therefore, to ensure high gradient values, the disc mass contribution to the mass function cannot be too small. An example of this interplay between various mass components will be discussed later in the example of galaxy NGC 4302 with its high measured gradient.

\subsection{Vertical gradient estimates in the global disk model}The gross dynamical mass distribution in a flattened galaxy may be approximately described by a thin disc, assuming axial symmetry and circular concentric orbits. In addition, it can be also assumed that the azimuthal component of the rotation above the galactic mid-plane can be determined by equating the radial component of the gravitational force to the centrifugal force (in cylindrical coordinates). This is called the quasi-circular orbits approximation. Validity of this approximation was substantiated in the case of the Milky Way Galaxy by analysing the motions of test bodies in the gravitational field of a thin disc \citep{bib:MNRASgradient}. In spite of its simplicity, the model turned out to work strikingly well for the gradient description in the Galaxy (this approximation was discussed in more detail in \citealt{bib:MNRASgradient}). We expect that this simplified description should be applicable in general for similar galaxies.

 In the disc model, the column mass density associated with the galactic mass distribution is represented by the equivalent surface mass density of a substitute thin disc, in accord with the following formula \citep{bib:jalocha_apj}
\[
\sigma(r)=
\frac{1}{\pi^2G} \mathcal{P} \left[\int\limits_0^r
v_{\sigma}^2(\chi)\biggl(\frac{K\br{\frac{\chi}{r}}}{ r\
\chi}-\frac{r}{\chi}
\frac{E\br{\frac{\chi}{r}}}{r^2-\chi^2}\biggr)\ud{\chi}+\dots
\right.\nonumber\]
 \begin{equation}  \label{eq:SigmafromRotCrv} \left.
\phantom{\sigma(r)= \frac{1}{\pi^2G} \mathcal{P} (\int\limits_0^r
v_{\sigma}^2(\chi)} \dots+
\int\limits_r^{\infty}v_{\sigma}^2(\chi)
\frac{E\br{\frac{r}{\chi}} }{\chi^2-r^2}\,\ud{\chi}\right].
\end{equation}
Here, the integration is understood in the principal value sense (both integrands are divergent, however the result of integration is finite). This substitute surface mass density can be used for an approximate description of the gravitational field produced by a flattened galaxy. This in turn, assuming the circular orbits approximation, provides us with a simple tool for predicting the azimuthal velocity component above the galactic mid-plane and, consequently, to determine the velocity fall-off rate as a function of the vertical distance from the mid-plane.

Given a surface mass density, the azimuthal component of the velocities of test particles in the gravitational field of a thin disc can be estimated in the circular orbits approximation, using the following expression \citep{bib:MNRASgradient}:
\[
{v_{\varphi}^2(r,z)}=\!\int\limits_0^{\infty}\!\!\!\frac{2\, G\sigma\br{\chi{}}\chi{}\ud{\chi{}}}{\sqrt{\br{r-\chi{}}^2+z^2}}\!\br{\!K\sq{X}-
\frac{\chi^2-r^2+z^2}{\br{r+\chi}^2+z^2}\,E\sq{X}\!}\!,\]
\begin{equation}\label{eq:VOverDiskFromSigma} X=-\sqrt{\frac{4r\chi{}}{\br{r-\chi{}}^2+z^2}}<0.\end{equation}
The model gradient value calculated as $\partial_z
{v_{\varphi}(r,z)}$, is a function of $r$ and $z$, whereas the gradient measurements are presented by giving a single number (and its error), being some average over an extended region. Therefore, to compare with, in a given region one could determine a mean
  value of $\partial_z
{v_{\varphi}(r,z)}$ over that region and the corresponding deviation
from the mean.

Other methods of vertical gradient estimation seem more relevant in the context of realistic gradient measurements, however. To mimic such measurements, one can specify in a given region an array of azimuthal velocities predicted by \eqref{eq:VOverDiskFromSigma}, and calculated for various pairs $(r,z)$. For each specified
variable $z$, separately, one can calculate the mean velocity by carrying out a summation over all radii in the region and then finding the slope of a linear fit to the mean-value data found at various $z$. We call this gradient estimation  the \textit{method-I}. Another estimation of the gradient value is obtained by finding the slope of a linear fit to the velocity values calculated for various $z$, separately at each fixed $r$, and then finding an average of the slopes taken over all radii in this region. This we call the \textit{method-II} gradient.
 When gradient $\partial_z
{v_{\varphi}(r,z)}$ is almost constant in a given region, both the methods should give results consistent  with each other. When $\partial_z
{v_{\varphi}(r,z)}$ is not constant in that region, the mean values over that region, determined based on both methods, could be still regarded as describing the observed fall-off rate well, provided the dispersion of the slopes discussed was comparably small.

 There are other, more or less complicated methods of gradient estimation possible, which we used in \citep{bib:MNRASgradient}, but all they led to similar results in the case of the Milky Way Galaxy. We expect that various method of gradient estimation should also give results consistent with each other for other spiral galaxies, therefore we decided here to use only the two methods described above.

\section{The results}
\subsection{NGC 2403}
\begin{figure}
   \centering
      \includegraphics[width=0.5\textwidth]{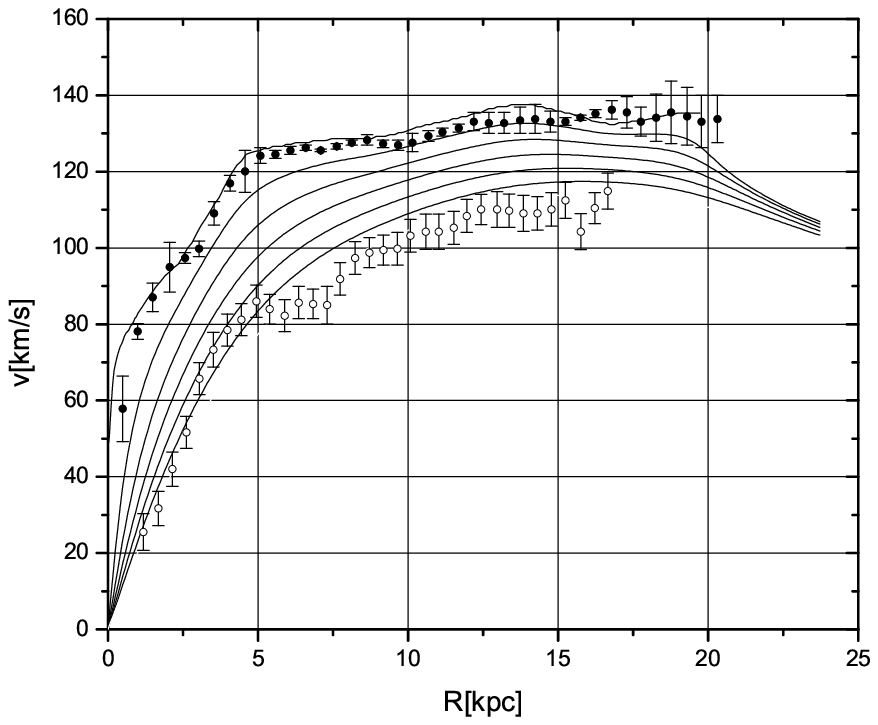}
      \caption{\label{fig:2403rot} Galaxy NGC2403: disc rotation curve (solid circles) and anomalous gas rotation curve above the disk (open circles), both from  \citet{bib:Fraternali2403}. The solid lines
      are rotation curves predicted by our model at various heights above the mid-plane
      at $z=0.6, 1.2, 1.8, 2.4\ \textrm{and}\ 3.0\,\kpc$. The topmost solid line is the \citet{bib:sofue}'s rotation curve.}
  \end{figure}

We use two rotation curves for galaxy NGC 2403, published in \citealt{bib:sofue} and \citealt{bib:Fraternali2403}. We assume the
distance of $3.24\,\mpc$, similarly as in \citealt{bib:sofue}, and find the integrated mass in disk model  to be
$5.22\times10^{10}\msun$.
 Unfortunately, owing to the insufficiently high inclination, it was impossible to determine precisely the vertical extent of the gas clouds, the velocity of which was measured above the cold disc \citealt{bib:Fraternali2403}. It was only possible to estimate the gas-layer thickness to be of about $3\,\kpc$ and the cold-disc thickness of about $0.4\,\kpc$
 \citep{bib:Fraternali2403}. Therefore, we assumed the region  $z\in(0.4,3)\,\kpc$ for the gradient estimation.
 In Fig. \ref{fig:2403rot}
 are shown the rotation curves from \citep{bib:sofue} and \citep{bib:Fraternali2403}, measured at the mid-plane. Both the curves are consistent with each other. Also shown is the rotation curve of the anomalous gas observed above the disc, somewhere in the interval mentioned. In between the curves are shown the rotation curves predicted in our model at various heights above the mid-plane. Our results may suggest that \citep{bib:Fraternali2403}'s rotation curve of the anomalous gas,  was measured even slightly above $3\,\kpc$.

We estimated the
gradient value using the method-II in the region $r\in(1,16.5)\,\kpc$ defined by the boundary points of the anomalous gas  measurements. We assume the radial step size of  $0.5\,\kpc$. The vertical interval is $z\in\br{0.4,2.4}\,\kpc$, with the step size of
$0.4\,\kpc$. We did not use the  method-I,  which might have led
to large errors owing to very extended radial region (method-I is best for the regions over the flat fragment of rotation curve, where the rotation above the disk is roughly independent of the radial variable).

\begin{figure}
   \centering
      \includegraphics[width=0.5\textwidth]{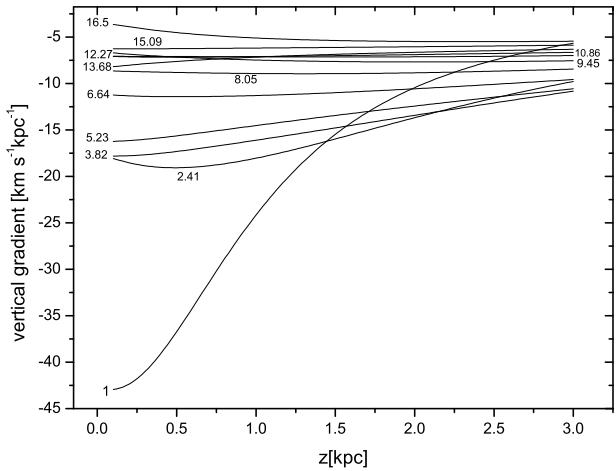}
      \caption{\label{fig:2403new} Galaxy NGC2403. Gradient values predicted in the global disc model at various heights above the mid-plane.}
  \end{figure}

The method-II gradient value is
$-10\pm4\gu$,
consistent with
\citet{bib:Fraternali2009}'s result
of about $-12\gu$ (the error was not published). Fig. \ref{fig:2403new}
illustrates the gradient behaviour
at various heights above the mid-plane.

\subsection{NGC 4559}
\begin{figure}
   \centering
      \includegraphics[width=0.5\textwidth]{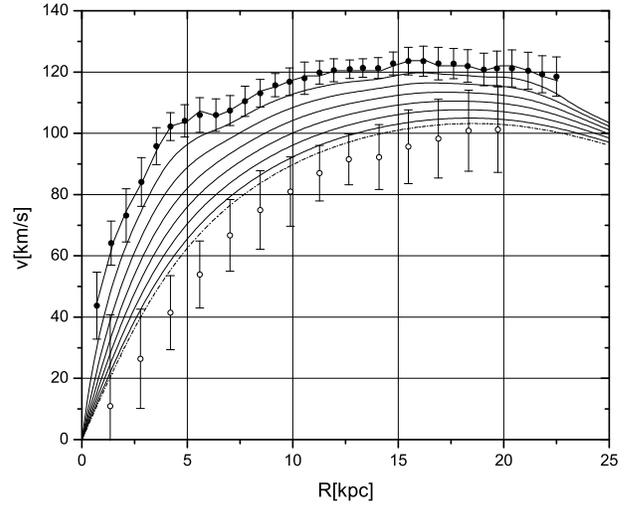}
      \caption{\label{fig:4559rot}Galaxy NGC 4559:
       disc rotation curve (solid circles) and the anomalous gas rotation curve above the disk (open circles), both from \citet{bib:Barbieri4559}.  The solid lines
      are rotation curves predicted by our model at various heights above the mid-plane
      at $z=0.6, 1.2, 1.8, 2.4, 3.0\ \textrm{and}\ 3.6\,\kpc$.
      The dashed line is the rotation curve predicted at $z= 4\,\kpc$.
       }
  \end{figure}

The vertical  gradient in this galaxy was observed  in the region $z\in(0.2,4)\,\kpc$ \citep{bib:Barbieri4559} and was determined in the same way as for galaxy NGC
2403. At the assumed distance of $9.7\,\mpc$ to this galaxy,  the integrated mass corresponding in our model to
\citet{bib:Barbieri4559}'s rotation curve,  is
$4.82\times10^{10}\msun$.
 In Fig.
\ref{fig:4559rot}
are shown the rotation curves of the cold disk of the anomalous  gas and the rotational velocities
predicted in our model at various heights above the mid-plane, out to $z=4\,\kpc$,
a  region in which the rotation of gas has been observed. As previously, the exact height of measurements is unknown but is located somewhere in the interval $z\in(0.2,4)\,\kpc$.
Our predictions are consistent with the measurements; in particular, the predicted rotation at $z=4.0\,\kpc$ overlaps within the error bars with the rotation curve of the anomalous gas. This may also suggest that the gas was measured slightly above $4.0\,\kpc$.
The method-II gradient was determined using the predicted rotation curves within the radial interval $r\in(1.3,19.7)\,\kpc$ using the same steps as for the rotation measurement points, and within the vertical interval  $z\in(0.2,3.7)\,\kpc$ with the assumed step size  of $0.7\,\kpc$.

\begin{figure}
   \centering
      \includegraphics[width=0.5\textwidth]{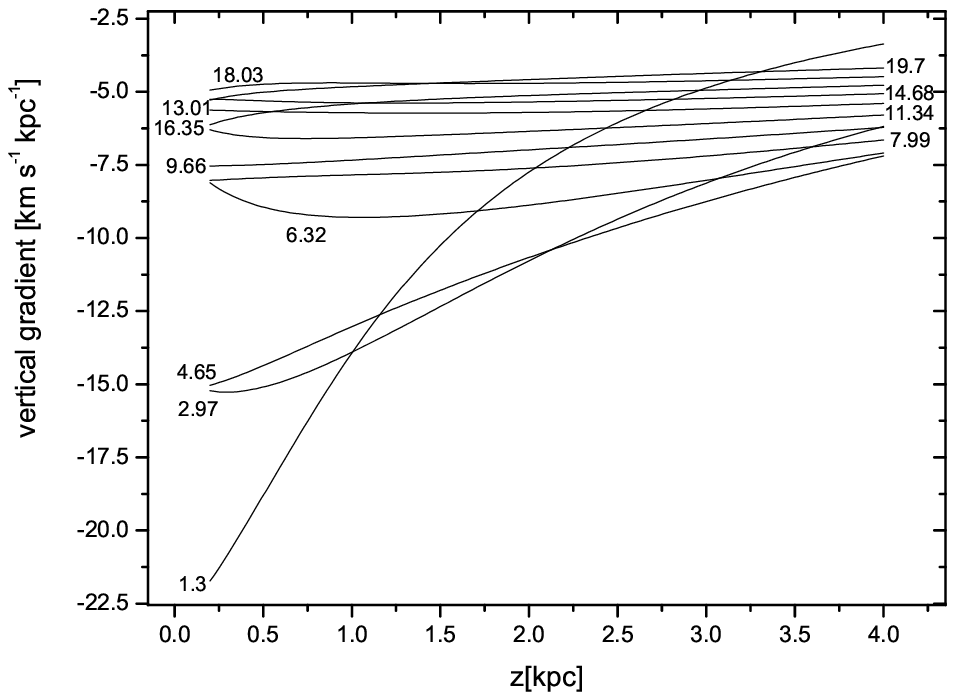}
      \caption{\label{fig:4559new} Galaxy NGC 4559: gradient values predicted in the global disk model at various heights above the mid-plane. }
  \end{figure}

Method-I is not used for the same reasons as for galaxy NGC
2403. The method-II gradient value is
$-7.2\pm2.4\gu$.
This result is lower than
that given by \citep{bib:Fraternali2009} of about $-10\gu$ (again, the error was not published). Since this result  must also have an error, we are justified to say that our prediction is consistent with gradient observations.
Fig. \ref{fig:4559new} shows the gradient behaviour
at various heights above the mid-plane.

\subsection{NGC 4302}
\begin{figure}
   \centering
      \includegraphics[width=0.5\textwidth]{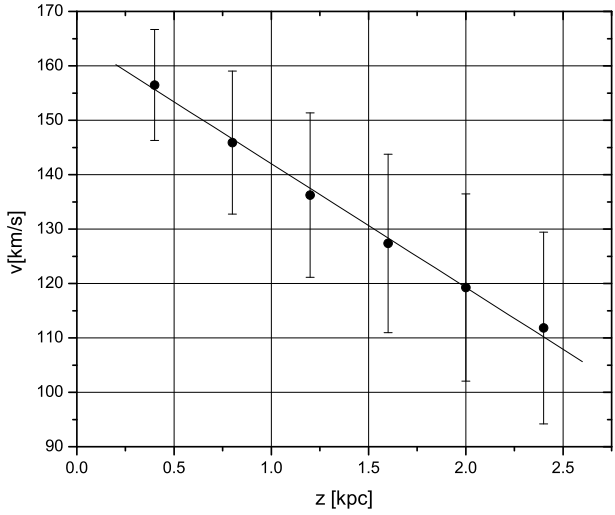}
      \caption{\label{fig:4302metoda1} Galaxy NGC 4302:
      azimuthal velocity as a function of the vertical distance from the mid-plane.
      The points represent the velocity values averaged over the interval $r\in(2.5,6)\,kpc$,
      the bars show the standard deviation in that interval. }
  \end{figure}

\begin{figure}
   \centering
      \includegraphics[width=0.5\textwidth]{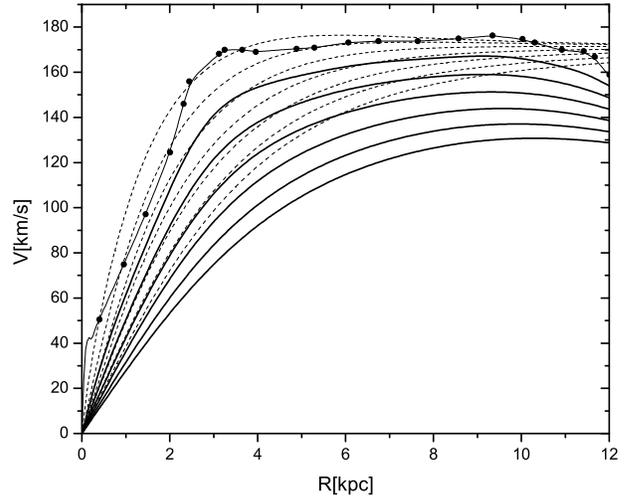}
      \caption{\label{fig:4302rot}
      The measured rotation curve of galaxy NGC 4302 (\citealt{bib:Heald4302}, solid circles), the rotational velocity
      above the mid-plane predicted in the  global disk model, shown at various heights above the mid-plane in the interval $z\in(0.6,3.6)\,\kpc$ in steps of $0.6\,\kpc$ (solid lines), and the rotational velocity above the mid-plane predicted by the two-component model presented in Fig.
      \ref{fig:dwiekomp}, shown in the same interval at the same heights above the mid-plane (dashed lines).}
  \end{figure}

\begin{figure}
   \centering
      \includegraphics[width=0.5\textwidth]{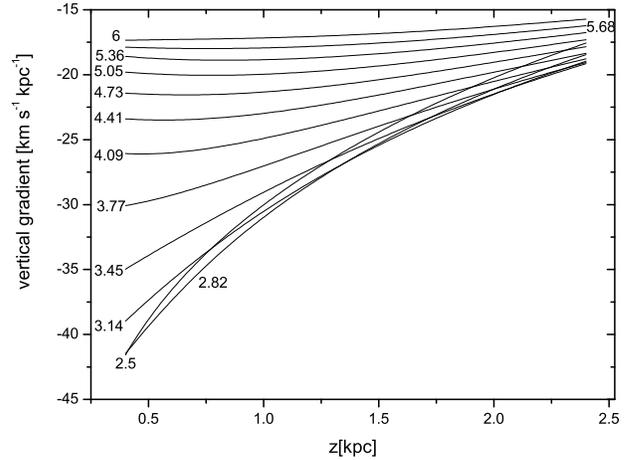}
      \caption{\label{fig:4302new} Galaxy NGC 4302: gradient values predicted in the global disc model at various heights above the mid-plane.  }
  \end{figure}

The rotation curve of galaxy NGC 4302 was taken from \citet{bib:Heald4302}.
The integrated mass at the assumed distance of $16.8\,\mpc$ in the disk model
is $5.05\times10^{10}\msun$. The galaxy has a large inclination; therefore the gradient measurement of the azimuthal velocity was made analogously to that for NGC 891 \citep{bib:Heald4302}.
\citet{bib:Heald4302} measured
 the gradient in the region $z\in(0.4,2.4)\,\kpc$, $r\in(2.5,6)\,\kpc$ and obtained the gradient value of about $-30\gu$ ($-31\pm19.8\gu$ in the south side). For the gradient calculation in the disk model, we assumed the same $z-r$ region as in \citet{bib:Heald4302}, with assumed $z$-step size of $0.4\,\kpc$ and $r$-step size of $0.5\,\kpc$.
  In figure \ref{fig:4302metoda1}
 the azimuthal velocity averaged over $r$ is shown as a function of the vertical distance.
 The method-I gradient is
  $-22.7\pm8.4\gu$ and the method-II gradient is
 $-22.3\pm4.0\gu$.
 These values overlap within the errors with those for \citet{bib:Heald4302}'s measurements.
 Fig. \ref{fig:4302rot} shows the rotational velocity above the mid-plane predicted in the global disc model and in a two-component model.
  Fig. \ref{fig:4302new}
  shows the gradient behaviour at various heights above the mid-plane.
  \begin{figure}
   \centering
      \includegraphics[width=0.5\textwidth]{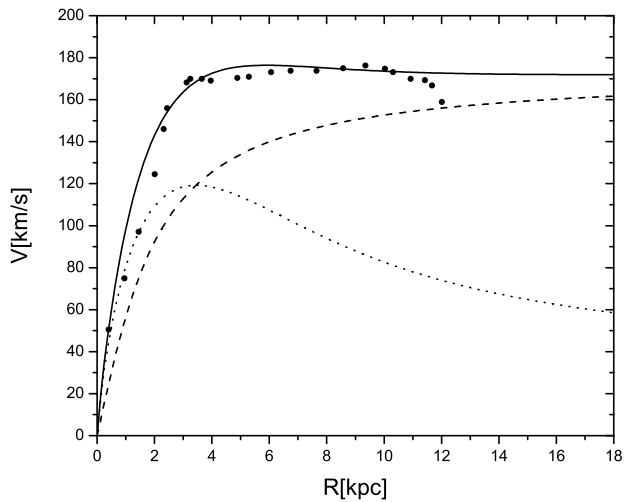}
      \caption{\label{fig:dwiekomp}{The least-squares fit of a two-component model of galaxy NGC 4302 considered in the text, fitted to the rotation curve (\citealt{bib:Heald4302}, solid line). The model rotation curve is decomposed to the exponential disc (dotted line) and to the dark
halo component (dashed line).}}
  \end{figure}
NGC 4302 illustrates well the influence of the spherical component on the gradient properties. To contrast this with the previous situation of purely a thin disc, consider a spherical halo with the density profile
$$\rho(r,z)=\frac{a_o b_o^2}{b_o^2+r^2+z^2}$$
comprising most of the galaxy mass, with the remaining mass attributed to an exponential thin disc. Now, the method-II gradient is strongly reduced to
$-14.8\pm4.1\gu$.
Fig. \ref{fig:dwiekomp} shows the least-squares fit of the two-component model.
Surely a mass model with arbitrarily assumed halo and disc-mass profile, cannot accurately account for the rotation curve. With a maximal halo model one can account for the rotation perfectly, but the corresponding method-II gradient value is even smaller,
$-11.1\pm4.5\gu$.
Accordingly, for mixed mass models with various relative mass profiles, the expected gradient value is between
$-11.1\gu$ for a more spherical-like mass distribution and
$-22.3\gu$ for a more disc-like mass distribution. To see better the influence of a heavy halo, let us estimate the gradient value in the mid-plane vicinity using the rotation data at points $r\in(3.5,7)\kpc$ in steps of $0.5\kpc$, and $z\in(0.06,1.06)\kpc$ in steps of $0.2\kpc$. The method-II gradient changes only a little in the disk model
$-21.1\pm6.0\gu$,
but for a "disk plus halo" model it reduces substantially, to
$-10.4\pm4.2\gu$. The greater the radial distance, the greater is the halo influence and the less important the disc component.  At small $z$, the gradient strongly decreases for the spherical distribution, whereas it is still high for the disc-like distribution. It is worth of noting the high gradient dispersion relative to the gradient value
$-10.4\pm4.2\gu$ in this region. The gradient varies more with $r$ in the spherical model than in the disk model and therefore with increasing halo and decreasing $z$ the gradient will be more dependent on $r$. The observations, however, do not confirm such a behaviour. Therefore, to see the gradient features better it would be worth repeating the rotation measurements in galaxy NGC 4302 at larger radii, closer to the mid-plane.

\subsection{NGC 5775}
\begin{figure}
   \centering
      \includegraphics[width=0.5\textwidth]{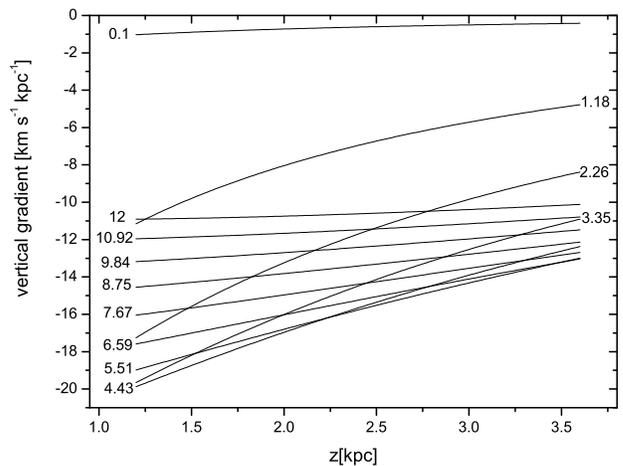}
      \caption{\label{fig:5775new} Galaxy NGC 5775: gradient values predicted in the global disk model at various heights above the mid-plane.   }
  \end{figure}

We use the rotation curve of galaxy NGC 5775 published in \citet{bib:Irwin5775}. The integrated mass is $9.65\times10^{10}\msun$ at the
assumed distance of $24.8\,\mpc$. The gradient measurements of the azimuthal
velocity  was carried out for this galaxy in \citet{bib:Heald5775} in a very extended region covering $r\in(0,12)\,\kpc$ and $z\in(1.2,3.6)\,\kpc$. This excludes the use of method I, since the azimuthal velocities at such a wide range of radii would differ too much, and the dispersion of the gradient estimation might be larger than the gradient value. Therefore, we
used only the method-II, which gives
 $-12.0\pm4.3\gu$.
The assumed  $r-z$ region is the same as that used for the measurements, and the assumed
steps are $\Delta{}z=0.6\,\kpc$ and $\Delta{r}=1.0\,\kpc$. The measured gradient value given in
\citet{bib:Heald5775} is $8\pm4\gu$ and is consistent within errors with our predicted value. Figure \ref{fig:5775new}
     shows the gradient behaviour
at various heights above the disk.
  \begin{figure}
   \centering
      \includegraphics[width=0.5\textwidth]{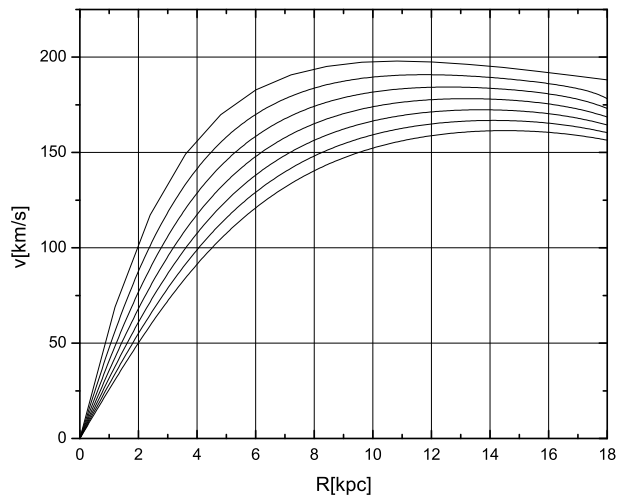}
      \caption{\label{fig:5775rot} Galaxy NGC 5775. From the top -- the rotation curve from \citet{bib:Irwin5775}, rotation curves predicted in disk model at various heights above the mid-plane for $z\in(0.6,3.6)\,\kpc$ in steps of $0.6\,kpc$. }
  \end{figure}
Fig. \ref{fig:5775rot} shows the rotational velocity above the mid-plane predicted in the disc model.

\section{Conclusions}
The vertical gradient in the rotational velocity of spiral galaxies can be satisfactorily explained in the framework of a global thin-disc model. The observed direction of the fall-off in rotational velocity (in the direction vertical to the galaxy mid-plane) and the high observed gradient values suggest that the gross mass distribution in spiral galaxies might be flattened, i.e. disc-like rather than spheroidal. With this hindsight, assuming that the disc comprises most of the dynamical mass inferred from the rotation curve, we calculated the vertical gradient of the azimuthal velocity in the quasi-circular orbits approximation.

  We examined four spiral galaxies, obtaining gradient values ranging from
  $-7.2\pm 2.4\gu$ to $-22.7\pm8.4\gu$.
We observed a general tendency in the gradient behaviour: for sufficiently large radii, the gradient values tend to be independent of the height above the galactic mid-plane.

Our predictions are consistent within errors with the measured gradient values. This agreement shows that the simple model of a global thin disc is sufficient to account for the measured, often high, gradient values. It is worth recalling that the global disc model gives the high gradient values required for galaxies like NGC 4302 or NGC 891 (see \citealt{bib:MNRASgradient}), which are difficult to explain by other models such as the ballistic model, the latter often predicting gradient values several times lower than those observed.
Depending on the parametrization, the ballistic model predicts the gradient values of $-8\gu$ for NGC 4302, but more realistic parameters give lower values of about $-1\gu$ \citep{bib:Heald4302},
much smaller than the measured ones, even reaching $-30\gu$, whereas the averaged rotation fall-off we obtain for this galaxy is
$-22.7\pm8.4\gu$.
For galaxy NGC, 5775 the ballistic model gives from $-2.4$ to $-4.3\gu$ \citep{bib:Heald5775},
again less than in our model (we obtain $-12.0\pm4.3\gu$), but, since the measured gradient is comparably low ($-8\pm4\gu$), the ballistic model prediction also overlaps with the measurement.
However, for galaxies with high gradient values our gradient modelling gives much better results than the ballistic model. We stress that here we simply compare, using the example of particular galaxies, the results of our model with the results of the ballistic model (with assumed parameters similar to those in the cited papers), and we do not make any assessment of the ballistic model with respect to the concepts underlying it.

It is important to stress that, based only on the mass distribution precisely related to a given rotation curve, the disc model, apart from accounting for high gradient values, gives the correct low gradient when the measured gradient is low. Of course, other models are also capable of giving correct results for galaxies with low gradients. For example, in the case of galaxy NGC 5775, the disc model predicts
    $-12.9\pm2.8\gu$, the maximal halo model
$-6.5\pm2.2\gu$,
and the ballistic model gives from $-2.4\gu$ to $-4.3\gu$. All of these values agree with the measurement $-8\pm4$. It is therefore important to have precise rotation measurements in the mid-plane vicinity at large radii, since they could discriminate between flattened and halo-dominated mass distributions. This is especially important for galaxy NGC 4302 with its high gradient.

The results of the current paper, and of our previous paper concerning the gradient study in the Milky Way Galaxy \citep{bib:MNRASgradient}, provide strong arguments that the gross mass distribution, at least in some spiral galaxies, might be flattened disc-like rather than spheroidal, unlike the suggestion of massive spheroidal dark halo models. Another conclusion is that the gas motion above the galactic discs of spiral galaxies might be governed by the gravitational potential of the galactic disc alone rather than by non-gravitational physics such as matter fountains.

\end{document}